\documentclass[12pt]{article}
\pdfoutput=1 

\usepackage[a4paper,margin=20mm]{geometry}

\usepackage[utf8]{inputenc}

\usepackage[super,comma,sort,compress]{natbib}
\usepackage{amsmath}
\usepackage[]{units}
\usepackage{gensymb}
\usepackage[font=small]{caption}
\usepackage{graphicx}
\usepackage{epstopdf}
\date{}

\title{\textbf{Ultrafast lithium diffusion in bilayer graphene}}

\author{M. Kühne$^1$, F. Paolucci$^{1,2}$, J. Popovic$^1$, P. M. Ostrovsky$^{1,3}$, J. Maier$^1$ \& J. H. Smet$^1$}

\makeatletter
\renewcommand\@biblabel[1]{#1.}
\makeatother

\begin{document}
\noindent
\Large\textbf{\textsf{Ultrafast lithium diffusion in bilayer graphene}}

\bigskip
\bigskip
\normalsize
\noindent
M. Kühne$^1$, F. Paolucci$^{1,2}$, J. Popovic$^1$, P. M. Ostrovsky$^{1,3}$, J. Maier$^1$ \& J. H. Smet$^1$

\bigskip
\bigskip
\noindent
$^1$Max Planck Institute for Solid State Research, 70569 Stuttgart, Germany

\medskip
\noindent
$^2$NEST, Istituto Nanoscienze-CNR and Scuola Normale Superiore, 56126 Pisa, Italy

\medskip
\noindent
$^3$L. D. Landau Institute for Theoretical Physics RAS, 119334 Moscow, Russia

\bigskip
\bigskip

\noindent
\textbf{Solid mixed conductors with significant ionic as well as electronic conduction play a pivotal role for mass transfer and storage as required in battery electrodes. Single-phase materials with simultaneously high electronic and ionic conductivity at room temperature are hard to come by and therefore multi-phase systems with separate ion and electron channels have been put forward instead. Here, we explore bilayer graphene as a true single phase mixed conductor and demonstrate ultrafast lithium diffusion exceeding diffusion in bulk graphite by an order of magnitude and even surpassing diffusion of sodium chloride in liquid water. To this end, an innovative electrochemical cell architecture has been developed where the redox-reaction forcing lithium intercalation is localized at a protrusion of the device only. Its remainder consists of pristine bilayer graphene unperturbed by an electrolyte. The geometry lends itself to the use of magnetotransport machinery known from mesoscopic low-dimensional physics. Time dependent Hall measurements across spatially displaced Hall probes deliver a direct view on the in-plane diffusion kinetics. The device layout with a perimeterial electrochemical cell is transferable to other 2D materials as well as thin films and may promote a paradigm shift on the use of electrolytes in on-chip experiments.}
\bigskip

The electrodes of state-of-the-art Li-ion batteries consist primarily of mixed conducting compounds, in which, by ambipolar motion, Li-ions diffuse coupled to electron transport. The higher both ionic and electronic conductivities of these electrodes are, the faster one may insert (and remove) Li-ions up to full storage capacity. However, room-temperature values for the chemical diffusion coefficient of Li, $D^{\delta}$, are typically rather low in common Li insertion compounds, with higher values potentially conceivable in artificially designed heterogeneous mixed conductors.\cite{Chen2016} In practical electrodes, which are composites containing the insertion compound (typically as a powder) and additives such as binders, Li diffusion is a complex process with $D^{\delta}$ being an effective parameter. For insertion compounds with a highly anisotropic ionic conductivity such as bulk graphitic carbons,\cite{Dresselhaus1981, Enoki2003} reported values spread over many orders of magnitude as $D^{\delta}\approx\unit[10^{-12}-10^{-5}]{cm^2/s}$,\cite{Yazami1983, Takami1995,Levi1997,Funabiki1998,Yu1999,Piao1999,Ong2002,Yang2004,Levi2005,Persson2010} partially also reflecting difficulties to reliably extract this parameter from electrochemical measurements.\cite{Yu1999,Yang2004,Levi2005} Nanoscale derivatives of bulk insertion compounds might therefore offer a better platform both to determine and to exploit their likely similar if not better mixed conducting properties. As such, graphene - a single sheet of carbon atoms - has been used in various electrodes as helpful current-collecting admixture, and a possibly increased capacity over graphite by Li adsorption on both sides is being discussed.\cite{Kaskhedikar2009, Lee2012} Here, we study bilayer graphene, which features a single van der Waals-gap between its two carbon sheets suitable for Li intercalation. With current nanofabrication methods and characterization tools, we strive to conceive single-crystal bilayer graphene structures suitable to directly probe the intrinsic Li-diffusion kinetics.
\medskip

To this end we realize a novel on-chip electrochemical device layout in which a gate voltage can be applied to a bilayer graphene flake across an electrolyte, a scheme successfully employed in recent years to both electrostatically tune the charge carrier density in two dimensional materials and surface layers,\cite{Fujimoto2013,Ueno2014} but also to drive electrochemical effects\cite{Jeong2013, Petach2014, Bao2014, Yu2015, Shi2015, Xiong2015}. The electrolyte, however, introduces additional disorder when applied directly on top of a sensitive surface,\cite{Gallagher2015, Browning2016, Ovchinnikov2016} and may thus shield or deteriorate the intrinsic properties of the latter. We therefore position a suited Li-ion conducting polymer electrolyte at one end of bilayer graphene, where once solidified it remains exempt from creep or wandering and endures temperature sweeps down to $\sim\unit[1.5]{K}$ while keeping its room-temperature properties.\cite{Nair2011, Gonnelli2015} As such it facilitates local intercalation and directed diffusion of Li along the device, a considerable part of which is deliberately uncovered from the electrolyte and thus even accessible to local probe techniques else disturbed by its presence.
\medskip

In this paper, we first present the device design and demonstrate the reversible intercalation of Li-ions during the gate-controlled electrochemical lithiation/delithiation of bilayer graphene devices. \textit{In situ} electronic transport measurements provide access to the charge carrier density induced by the charge transfer from the intercalant. We provide corroborating evidence for the reversible intercalation of Li ions from their strong impact on the intervalley scattering rate of charge carriers, revealed in low-temperature magneto-transport measurements. After successful lithiation open bilayer graphene edges appear decorated with Li-rich material, most likely due to the reaction of intercalated Li with residues in the high vacuum atmosphere. Particular device designs exploiting the possible implications of this finding allow us to unambiguously proof the intercalation and diffusion mechanisms with Li residing between graphene layers only. We thus implement the most elementary unit of a graphite intercalation compound, with time-resolved \textit{in-situ} electronic transport measurements providing unprecedented direct access to the intercalate diffusion kinetics. We extract record in-plane chemical diffusion coefficients for Li at room temperature, up to $D^\delta=7\cdot\unit[10^{-5}]{cm^2/s}$, beyond the upper limit of values reported from studies on bulk graphite\cite{Yazami1983, Takami1995,Levi1997,Funabiki1998,Yu1999,Piao1999,Ong2002,Yang2004,Levi2005,Persson2010}. Moreover, this value, to our knowledge, is the highest chemical diffusion coefficient ever observed at room-temperature, only surpassed by motion along heterophase boundaries.\cite{Chen2016}
\bigskip

\noindent
\textbf{Device design for electrochemical lithiation}

\bigskip
\begin{figure}
\centering
\includegraphics{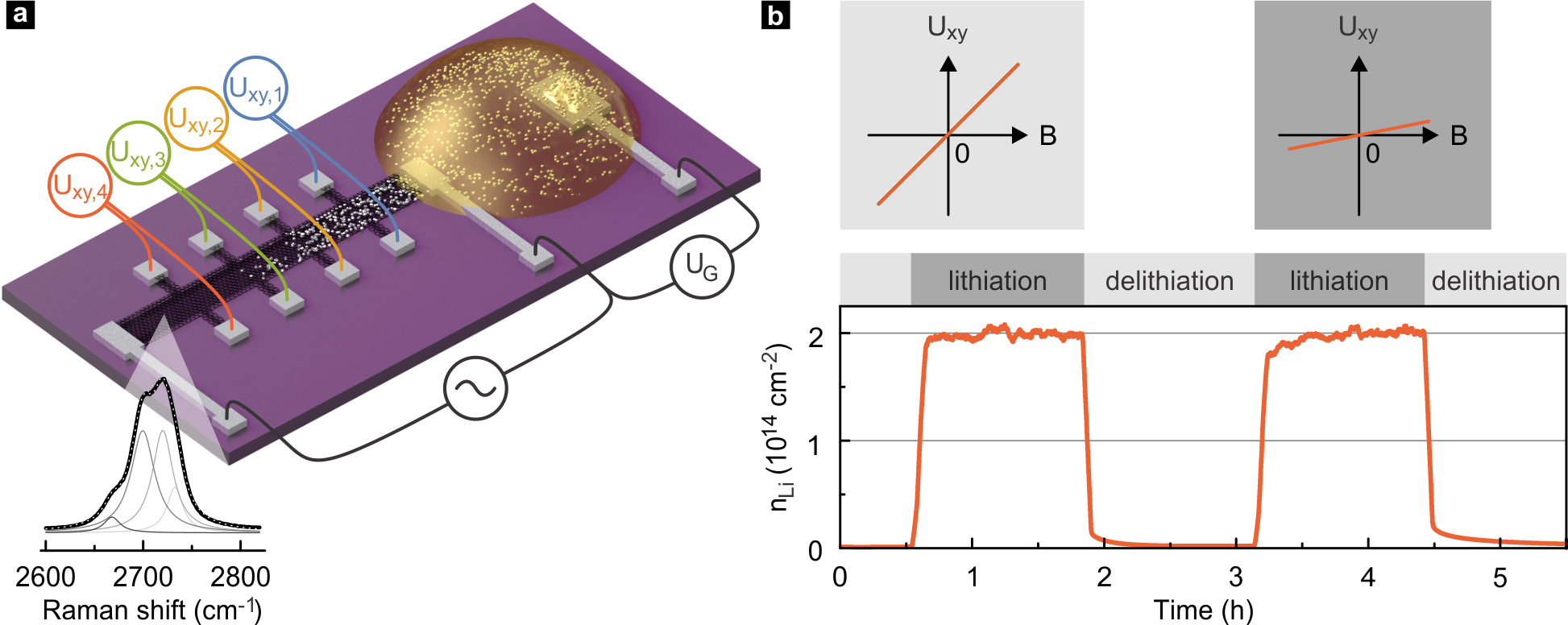}
\caption{Electrochemical device setup. (a) Schematic of the device (not to scale). Bilayer graphene etched into a Hall bar geometry. Electrodes either enable electronic transport measurements or serve as a counter electrode to control the lithiation. The inset shows the characteristic 4-component Raman scattering response of bilayer graphene (intensity in arb. units),\cite{Ferrari2006} measured for an  excitation wavelength $\lambda=\unit[488]{nm}$. (b) Reversible electrochemical lithiation of a bilayer graphene device, measured in the region not covered by the electrolyte. Lithiation (delithiation) at $U_G=\unit[0.05]{\text{V vs. Li/Li}^+}$ ($U_G=\unit[1.5]{\text{V vs. Li/Li}^+}$) induces a measurable increase (decrease) of the electron concentration as extracted from Hall measurements. The electron concentration reflects the amount of intercalated lithium $n_{\text{Li}}$. The Hall measurement was performed at $\unit[10]{T}$, in high vacuum and at room temperature.}
\label{fig1}
\end{figure}

\noindent
Fig.~1a schematically illustrates the design of our electrochemical setup. A mechanically exfoliated bilayer graphene flake is deposited on a SiO$_2$-terminated Si substrate and patterned into a Hall bar geometry. The bilayer nature is verified with Raman spectroscopy\cite{Ferrari2006} as shown in the inset to Fig.~1a. The patterned flake is fitted with Ti contacts to allow for longitudinal and Hall resistance measurements. An additional Ti pad, placed at one end but isolated from the flake, is created in the same step. Metallic Li is deposited on it inside an Ar-filled glovebox, thereby forming the counter electrode (CE). Finally, a polymer electrolyte drop is positioned at the periphery of the flake, covering only a small portion of the latter as well as the lithium CE. Subsequently, the polymer electrolyte is solidified by UV-curing in order to improve its mechanical stability and simplify device handling. Further details related to device fabrication and the electrolyte are deferred to the methods section.
\medskip

LiC$_6$ is the stoichiometry with the densest packing of Li achievable under normal conditions in bulk graphite samples.\cite{Winter2011} In the relevant case of Li storage in between graphene sheets only, bilayer graphene may be intercalated up to a stoichiometry of C$_6$LiC$_6$.\cite{Sugawara2011} The electrochemical setup then translates to the following galvanic cell: Li$^\circ$\textbar polymer electrolyte for Li-ion transport\textbar C$_6$Li$_y$C$_6$, where $y$ denotes the lithium content in the bilayer graphene host with $0\le y\le 1$. Here, Li metal (denoted as Li$^\circ$) provides a source or a sink for Li-ions and it serves as the reference potential. Of course, both Li$^\circ$ and C$_6$Li$_y$C$_6$ are electronic conductors, while the polymer electrolyte is an electronic insulator. The electrical potential of the C$_6$Li$_y$C$_6$ working electrode (WE) with respect to the Li metal counter electrode (CE) is related to the state of lithiation of the former. This potential varies to a great extent in a reversible manner during charging and discharging of the cell, as Li-ions intercalate into and deintercalate from the WE. The intercalation of a single Li-ion is accompanied by the charge transfer on the order of one electron (e$^-$) to the electronic states of the graphitic host, leaving the Li 2s level unoccupied.\cite{Enoki2003} This valence charge acts to screen the intercalant ion and is thus rather locally distributed.\cite{Holzwarth1992, Winter2003} There are contradictory reports concerning the extent of transferred charge in bulk Li$_y$C$_6$,\cite{Dresselhaus1981, Holzwarth1992, Winter2003} which is possibly dependent on the intercalant concentration. In the case of C$_6$Li$_y$C$_6$, recent \textit{ab initio} calculations favor a Li density-independent charge transfer on the order of $\sim\unit[0.88]{e^-}$ per Li atom.\cite{Guzman2014, Shirodkar2016} This is supported by experimental evidence for the full ionization of Li in C$_6$LiC$_6$ on SiC.\cite{Sugawara2011} Assuming a complete charge transfer, the induced electron density in C$_{6}$LiC$_{6}$ should amount to $n\approx\unit[6.36\cdot 10^{14}]{cm^{-2}}$, which agrees with a reported calculation of $\unit[6.1\cdot 10^{14}]{cm^{-1}}$.\cite{Kaloni2012} In such a highly lithiated state, the chemical potential of Li in C$_6$LiC$_6$ is only $E_{\text{chem}}^{y\rightarrow1}\approx\unit[0.15]{\text{V vs. Li/Li}^+}$ (inferred from values reported for bulk LiC$_{12}$, see, e.g., ref. \cite{Winter2011}), i.e., very close to the one of metallic Li. In the delithiated state, $E_{\text{chem}}^{y\rightarrow0}>\unit[1.5]{\text{V vs. Li/Li}^+}$. Hence, it should be possible to lithiate and delithiate bilayer graphene at a low and high potential vs. Li/Li$^+$, respectively.
\medskip

In the uncovered part of the bilayer graphene device, we anticipate that diffusive forces continue to drive intercalation outside and  far away from the electrolyte. Contrary to what occurs within the electrolyte from all perimeterial edges and defects within the bilayer, the diffusion outside of the electrolyte is well directed. The elongated geometry in conjunction with the spatially confined electrolyte ensures the alignment of the lithium density gradient with the extended dimension of the flake ideal for well-defined in-plane diffusion studies. The placement of a liquid or solid on top of bilayer graphene is highly invasive\cite{Browning2016} and produces large strain fluctuations\cite{Couto2014} and lattice distortions with a detrimental impact on its properties. In this device architecture with a perimeterial electrochemical cell only, these issues are avoided and the intrinsic diffusion coefficient can be measured. The increase of the Li-ion density during intercalation will be reflected in the charge carrier density of the bilayer. The latter can be monitored continuously as a function of time by measuring the Hall resistance across potential probes located on opposite edges of the sample, see Fig.~1a. To this end, we impose a constant dc-current through the bilayer flake and apply a constant perpendicular magnetic field $B=\unit[10]{T}$. The Hall voltage $U_{xy}(x)$ building up at different positions, $x = x_{i}$ with $i = 0,1,2,\ldots $, is directly related to the local charge carrier density $n(x)=BI/eU_{xy}(x)$.  Hence,  assuming a complete charge transfer of $1 e$ per intercalated ${\rm Li}^{+}$ with $e$ the elementary charge of an electron, the measured density of electrons in bilayer graphene, $n_{e}$, equals the density of intercalated lithium $n_{\text{Li}}$, minus an initial charge density from intrinsic doping sources $n_{\text{imp}}$. The latter can be extracted from a Hall measurement prior to Li intercalation. It follows that $n_{\text{Li}}(x,t)=n_{e}(x,t)-n_{\text{imp}}(x,t=0)$. Upon initiating delithiation, the carrier density should return to its original value. An example of the extracted charge carrier density during repeated intercalation (deintercalation) at $U_G=\unit[0.05]{\text{V vs. Li/Li}^+}$ ($U_G=\unit[1.5]{\text{V vs. Li/Li}^+}$) is illustrated in Fig.~1b. Indeed, significant changes in the Li density outside of the electrolyte-covered bilayer area can be deduced from these measurements, despite the large spatial separation between the Hall probes and the polymer electrolyte. In this sample, the electron density reaches a maximum value of approximately $2 \cdot 10^{14}\ {\rm cm}^{-2}$. This can be considered as a lower limit for the lithium density $n_{\text{Li}}$ in case the charge transfer ratio is less than the assumed value of one. Even though it is plausible that lithium is intercalated in between the two carbon sheets of bilayer graphene, it is challenging with the Hall method to distinguish whether Li-ions are located underneath, above or in between the two graphene layers. Before addressing the diffusion dynamics in more detail, low temperature magnetotransport and ex-situ characterization techniques are deployed to substantiate the claim that Li-ions are present in the bilayer region not covered by the electrolyte and that these ions have intercalated in between the graphene planes only.

\bigskip

\noindent
\textbf{Li intercalation probed by intervalley scattering}

\bigskip

\begin{figure}
	\centering
	\includegraphics{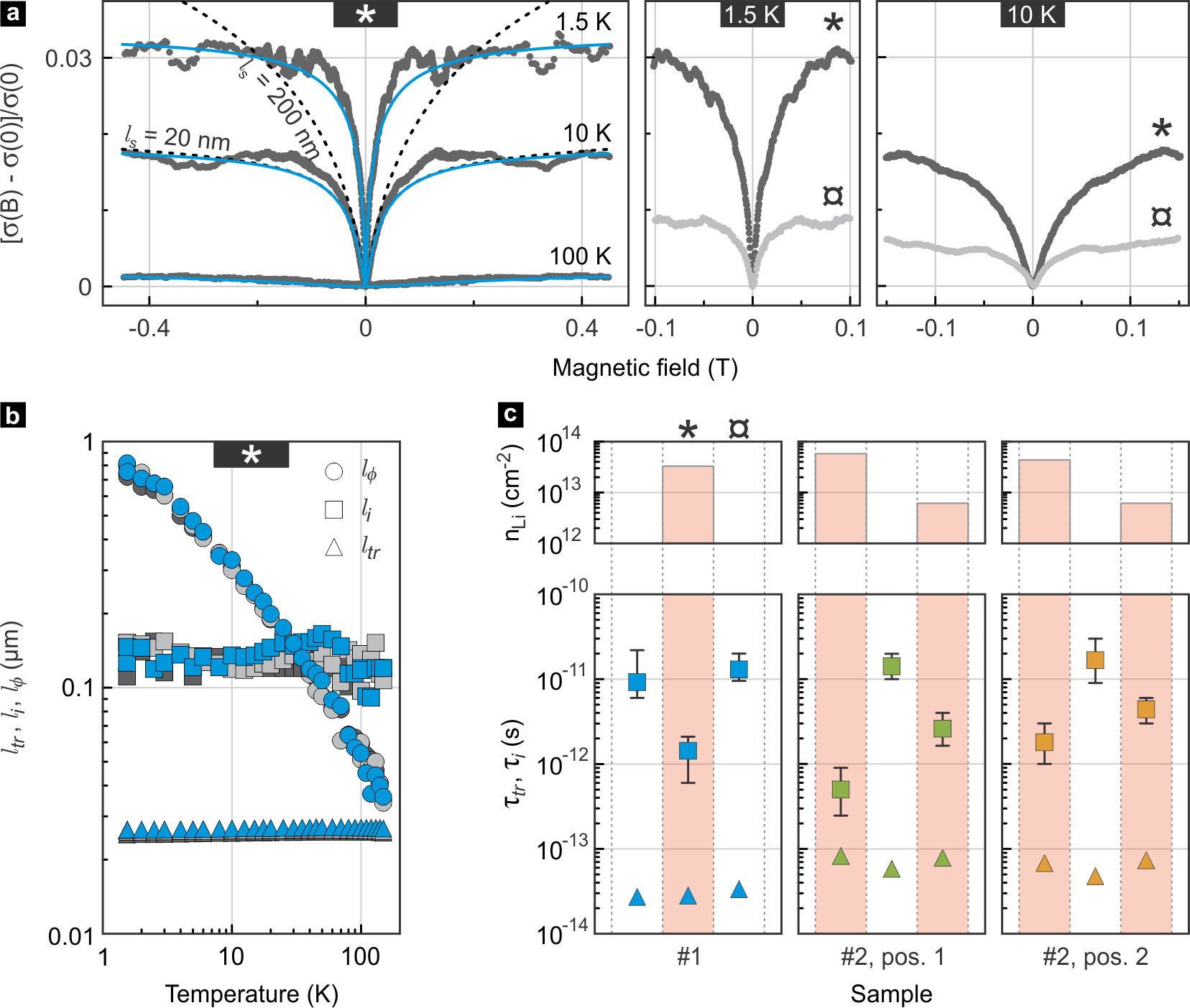}
	\caption{Weak localization in bilayer graphene devices at different states of lithiation. (a) Right panels: Exemplary magnetoconductivity traces measured on sample \#1 in a lithiated (dark grey dots) and delithiated (bright grey dots) state at different temperatures. Left panel: Weak localization traces in the lithiated state compared with Eq.~\ref{eq:WL}. Blue, solid lines are fits to the data using the first two terms in Eq.~\ref{eq:WL}. Black, dashed lines are plots of Eq.~\ref{eq:WL} with parameters as for the simplified fit at $\unit[10]{K}$ and two values for $l_s$ as indicated. (b) Temperature dependence of the phase relaxation length $l_\phi$ (circles), the intervalley scattering length $l_i$ (squares) and the mean free path $l_{tr}$ (triangles) for sample \#1 in its most lithiated state. (c) Intervalley scattering time (squares) and transport time (triangles) for sample \#1 and two different contact pairs of sample \#2. Columns are successive states of lithiation with the density $n_{\text{Li}}$ of intercalated Li shown in the column chart (top). White columns refer to pristine or delithiated states.}
	\label{fig:WL}
\end{figure}

\noindent
Li-ions inserted in between the graphene planes represent additional atomic-scale scattering sites for the electronic charge carriers and, hence, we may anticipate enhanced elastic scattering of the latter upon lithiation. Since their size is smaller than the lattice spacing, Li-ions enable the large momentum transfer required for scattering electronic charge carriers between the two valleys in bilayer graphene. A significant rise in the associated intervalley scattering rate $\tau_i^{-1}$ and a subsequent drop back towards its original value during a lithiation/delithiation cycle would be another striking confirmation for Li diffusion from the electrolyte at the device periphery into and out of the uncovered bilayer graphene.
\medskip

It is a fortunate coincidence that the weak localization effect,\cite{Altshuler1980} a magnetic field dependent quantum interference correction to the electron conductivity at low temperature, offers a powerful framework to detect changes in the intervalley scattering rate in the case of bilayer graphene, despite the elastic nature of the scattering process.\cite{Gorbachev2007,Kechedzhi2007} Unlike in conventional diffusive metals where only inelastic scattering affects weak localization due to decoherence, elastic scattering mechanisms involving the valley symmetry also generate quantum corrections to the conductivity in bilayer graphene, because of the intimate connection between the valley degree of freedom and the chirality of charge carriers in bilayer graphene and the obvious relevance of chirality for quantum interference effects of particle trajectories. The weak localization contribution, $\Delta \sigma$, to the magnetoconductivity in bilayer graphene is described by the following relation\cite{Gorbachev2007,Kechedzhi2007}
\begin{equation}
\label{eq:WL}
\Delta\sigma(B)=\frac{e^2}{\pi h} \left[F\left(\frac{B}{B_\phi}\right)-F\left(\frac{B}{B_\phi+2B_i}\right)+2F\left(\frac{B}{B_\phi+B_i+B_s}\right)\right],
\end{equation}
with
\begin{align*}
F(z)&=ln(z)+\psi\left(\frac{1}{2}+\frac{1}{z}\right)&\text{and}&&B_{\phi,i,s}^{\vphantom{-2}}&=\frac{\hbar}{4e}\cdot \frac{1}{l_{\phi,i,s}^{2}} = \frac{\hbar}{4e}\cdot \frac{1}{D \tau^{\vphantom{2}}_{\phi,i,s}}.
\end{align*}
Here, $\psi(x)$ is the digamma function and $l_\phi$, $l_i$, $l_s$ are the characteristic length scales for phase coherence, intervalley scattering and intravalley chiral symmetry breaking. While $l_i$ is dominated by sharp impurities (such as Li-ions), the length $l_s$ is limited by the trigonal warping of the Fermi surface as well as electron-hole puddles, ripples, dislocations and other smooth defects. The corresponding time scales follow from the charge carrier diffusion coefficient $D=v_F^2\tau^{\vphantom{2}}_{tr}/2$ as $l_{\phi,i}=\sqrt{D\tau_{\phi,i}}$. The transport time $\tau_{tr}=\sigma_0 m^\ast/(n e^2)$ is extracted from the zero field conductivity $\sigma_0(T) \approx \sigma(B=0,T) - \Delta\sigma(B,T)$ without quantum corrections. Both the Fermi velocity $v_F$ and the effective mass $m^\ast$ depend on the charge carrier density $n$ and are calculated using a simple tight-binding model of the bilayer graphene electronic structure.\cite{McCann2013} In all cases we confirm that $k_Fl_{tr}\gg1$ (with the Fermi wavevector $k_F$), i.e., charge carrier transport is in the diffusive, weakly disordered regime and Eq.~\ref{eq:WL} applies.
\medskip

The right panels in Fig.~2a illustrate the enhancement of the weak localization correction to the conductivity in the lithiated state (dark grey) compared with the delithiated state (light grey) of the sample at a temperature of 1.5 and 10 K. Successful fits to the data are typically obtained using only the first two terms in Eq.~\ref{eq:WL}, as $l_s$ is by far the shortest length. The good agreement between the two term model and our measurements is demonstrated in the left panel of Fig.~2a for some exemplary traces measured on a lithiated sample with $n_{\rm Li}\approx\unit[3.5\cdot10^{13}]{cm^{-2}}$. The grey dotted lines represent experimental data recorded at three different temperatures. The blue, solid lines are best fits to the data considering only the first two terms in Eq.~1. For the data measured at $\unit[10]{K}$, we additionally plot two traces for the full model (dashed lines) for two selected values of $l_s$ (20 nm and 200 nm). The 3rd term of Eq.~\ref{eq:WL} can indeed be neglected and $l_s$ is on the order of the transport mean free path, $l_{tr} = v_{F} \tau_{tr} \approx \unit[26]{nm}$, essentially set by the scale of the electron and hole puddles.\cite{Martin2008} In Fig.~2b we plot the temperature dependence of the characteristic length scales in addition to the mean free path $l_{tr}$. While both $l_i$ and $l_{tr}$ show no clear temperature dependence, $l_\phi$ drops with increasing temperature as dephasing mechanisms gain efficiency, thereby suppressing interference effects. In Fig.~2c we have extracted the intervalley scattering time $\tau_i$ on contact pairs from two different bilayer graphene samples and for different states of lithiation. We use the maximum spread in values obtained for $l_i$ from fits at temperatures between 1.4 and 10 K to determine the error bars shown in the plots. For comparison, we again include the transport times $\tau_{tr}$. While the latter vary from sample to sample (depending on quality and residual doping) and show no conclusive dependence on the state of lithiation, $\tau_i$ is significantly reduced in the presence of intercalated Li, i.e., the intervalley scattering rate $\tau_i^{-1}$ is enhanced. Not only does $\tau_i^{-1}$ increase upon Li intercalation, it also consistently decreases upon deintercalation, revealing the degree of reversibility of the intercalation process. Hence, there can be no doubt that Li is inserted and largely taken out again. Upon insertion, it assists in scattering charge carriers between the two valleys in bilayer graphene, not unlike what has been reported for In adatoms on single layer graphene.\cite{Chandni2015}

\bigskip

\noindent
\textbf{Ex-situ characterization after prolonged lithiation}

\bigskip

\begin{figure}
	\centering
	\includegraphics{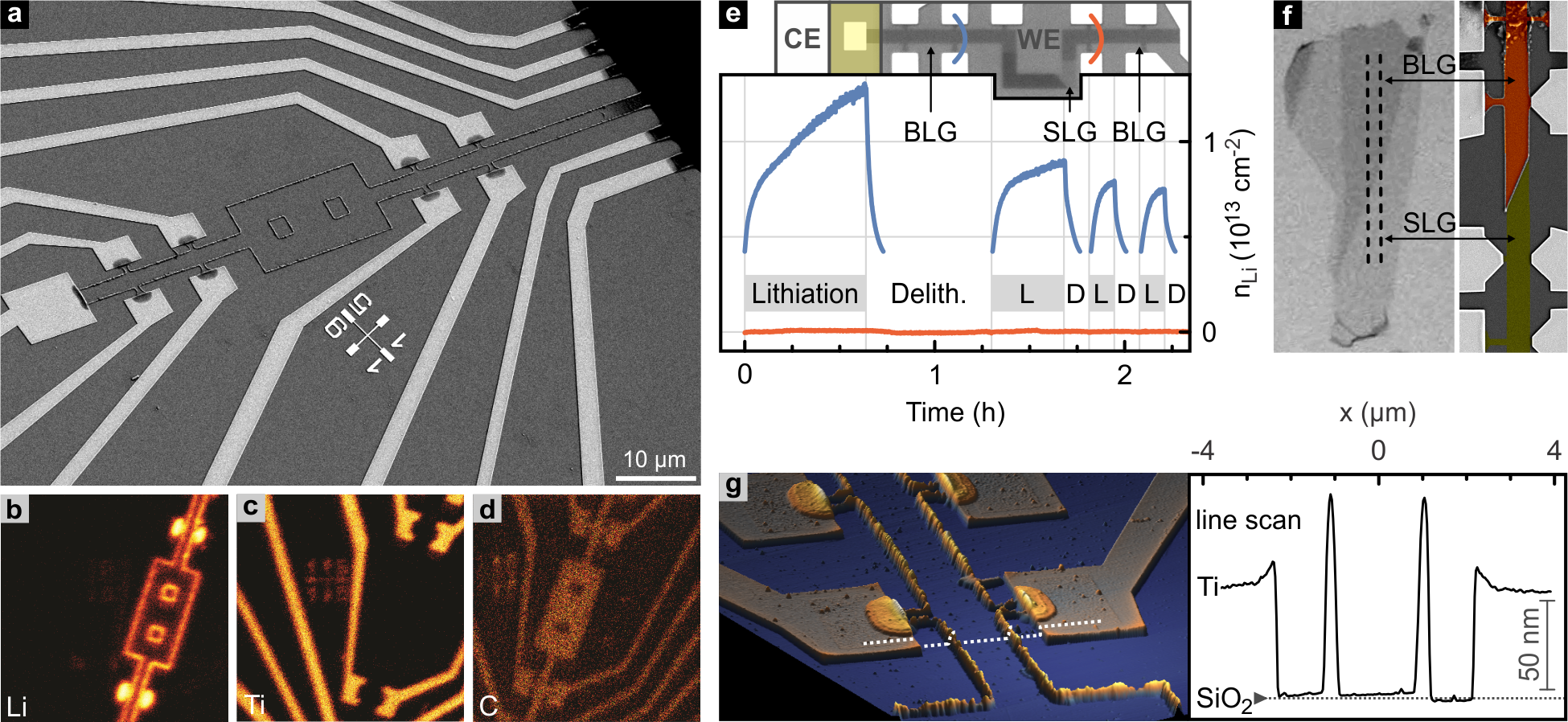}
	\caption{Revealing Li intercalation in between graphene sheets only. (a) Scanning electron micrograph of a bilayer graphene device with two $\approx\unit[6]{\mu m^2}$ etched holes after prolonged lithiation. (b-d) Spatially resolved ToF-SIMS of the central region in (a) for the Li, C and Ti channel. (g) Atomic force micrograph of a part of the same sample with a line scan acquired along the dashed line. (f) Optical micrograph (left, prior to etching and contact deposition) and scanning electron micrograph (right) of a second device with a natural bilayer (orange shaded area)/single layer (yellow shaded area) junction after prolonged lithiation. (e) Time-dependent lithium concentration $n_{\rm Li}$ derived from \textit{in-situ} measurements of the Hall voltage during lithiation/delithiation of a bilayer graphene device both before and beyond a natural single layer graphene junction. The schematic of the Li$^\circ$ (CE)\textbar polymer electrolyte (yellow shaded area)\textbar C$_6$Li$_y$C$_6$ (WE) cell includes an optical micrograph of the uncovered bilayer graphene area with probes used for the Hall measurements. The Hall probe pairs are marked with the same color as the corresponding data sets.}
	\label{fig:exsitu}
\end{figure}

\noindent
Ex-situ characterization of bilayer graphene devices after prolonged lithiation with scanning electron microscopy (SEM), atomic force microscopy (AFM) and time-of-flight secondary ion mass spectrometry (ToF-SIMS) allows us to collect evidence that Li intercalation only occurs between the two carbon planes. Intercalation in between the SiO$_2$ and graphene as well as Li migration on top of the graphene planes is not observed under our experimental conditions.
\medskip

Upon successful Li intercalation, bilayer graphene devices reveal decorated edges as unveiled in the SEM and AFM images of Fig.~3. This decoration is found to grow only during prolonged lithiation with high density of intercalated Li. With the lithium activity in Li-rich graphitic carbon being comparable to the one of metallic Li,\cite{Winter2011} we attribute the observed edge decoration to the high reactivity of the intercalated Li. Reaction partners may be provided either from the residual gas atmosphere (although experiments are carefully performed at $p\le\unit[1\cdot10^{-6}]{mbar}$) or from mobile hydrocarbons on the sample surface stemming from processing steps during device preparation or from the polymer electrolyte which possesses compounds with non-negligible vapor pressure. As intercalated Li may not leave bilayer graphene other than at edges or at suitable defect sites, immobile reaction products may agglomerate at such sites.
\medskip

ToF-SIMS reveals the high Li content of the edge decoration (Fig.~3b). A scanning electron micrograph of the same part of the sample is shown in Fig.~3a. There is a clear correspondence between the spatial distribution of Li in the ToF-SIMS data and the edges of the bilayer (compare with the C channel, Fig.3c), as well as the one of Ti (Fig.~3d) and the electrodes. Two quadratic holes of $\approx\unit[6]{\mu m^2}$ were etched on purpose into this bilayer before lithiation to corroborate that Li transport is not limited to the sample perimeter. It takes place between the graphene planes, since the edges of the holes get decorated the same way as the outer edges. The height signal of an atomic force microscopy (AFM) scan acquired on the same sample is displayed in Fig.~3g, together with the profile extracted along the indicated line crossing both the bilayer graphene and the Ti electrodes. It can be seen, that here both width and height of the edge decoration exceed $\unit[100]{nm}$. Such edge decoration stops at the natural junction with single layer graphene as imposingly revealed in the device of Fig.~3f, which consists of a bilayer section followed by a single layer region. It proves unequivocally intercalation of Li between the two carbon planes of bilayer graphene only, whereas intercalation between the SiO$_2$-substrate as well as Li migration on top of graphene planes can be excluded. Additional support for these assertions has been collected from in-situ time-resolved Hall measurements on a device where the bilayer is interrupted with a single layer graphene region (Fig.~3e), but continues afterwards. While the Hall signal changes due to Li diffusion up to the single layer, no change is observed in the bilayer part beyond the single layer portion. Our finding agrees with \textit{ab initio} calculations according to which Li intercalates in between graphene sheets only and does not reside on its surface (down to the single layer limit) in the absence of defects.\cite{Lee2012} Notwithstanding the fact that Li diffusion on the graphite surface was demonstrated, intercalation is energetically much more favorable.\cite{Mandeltort2012}
\medskip

The edge decoration is specific to the geometry addressed in this work with a spatial separation between the electrolyte and an uncovered region of the 2D material under study. Although the exact chemical nature of the edge decoration remains to be understood, similar processes must be expected in general for the intercalation of reactive species into layered materials with open edges or defects. These sites may act like a sink for the intercalant, which is undesirable when targeting high intercalant densities. It may be alleviated by optimizing the aspect ratio of the device to minimize its perimeter and likely also by covering or sealing the edges in a suitable manner.

\bigskip

\noindent
\textbf{Diffusion coefficient}

\bigskip

\begin{figure}
	\centering
	\includegraphics{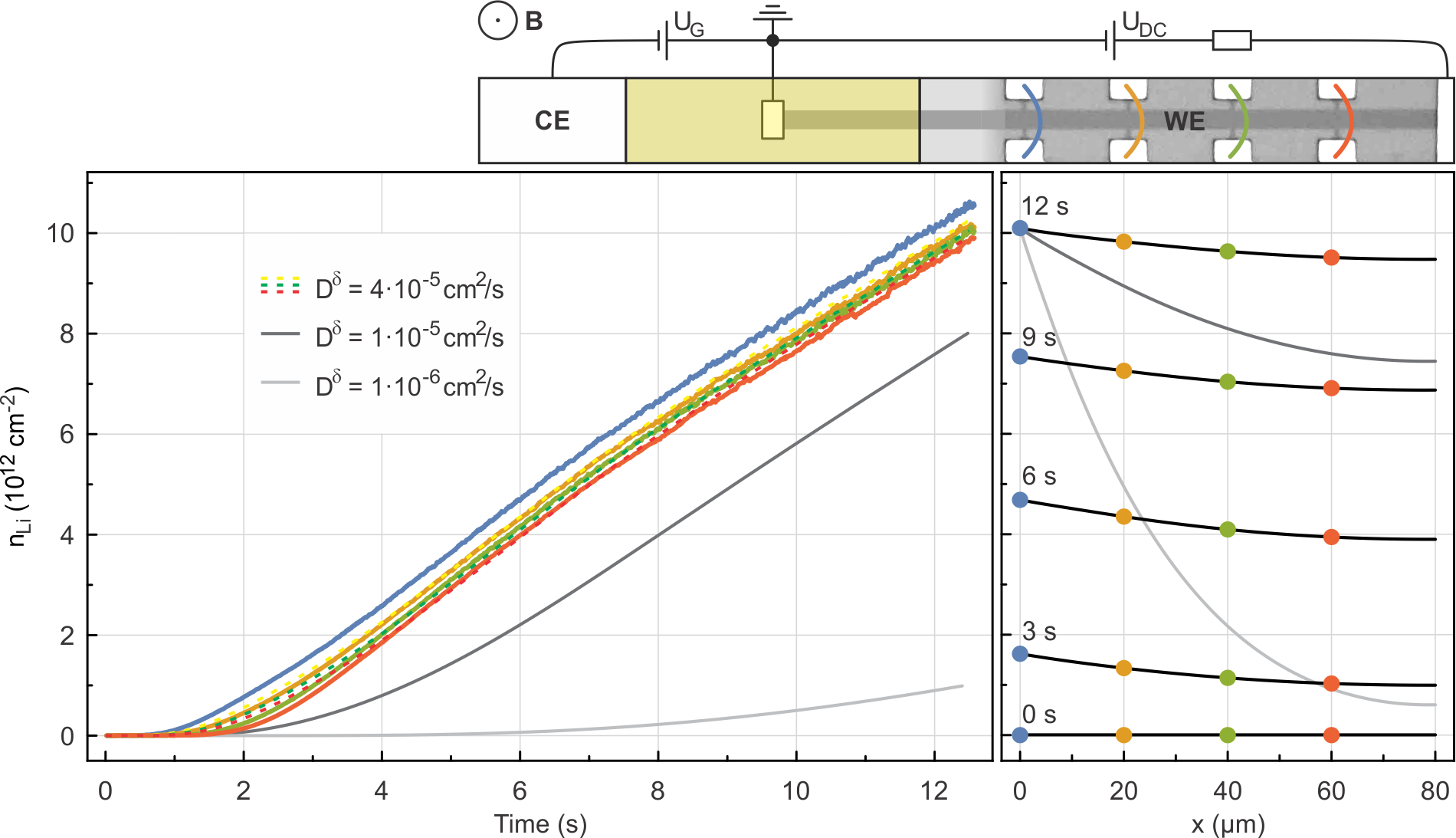}
	\caption{Direct measurement of Li diffusion. The schematic of the Li$^\circ$ (CE)\textbar polymer electrolyte (yellow shaded area)\textbar C$_6$Li$_y$C$_6$ (WE) cell includes an optical micrograph of the uncovered bilayer graphene area with probes used for the Hall measurements. Hall probe pairs are marked in the same color as the corresponding data set. Left panel: Li concentration n$_{\text{Li}}$ extracted from Hall measurements at $B=\unit[10]{T}$ and $T=\unit[300]{K}$ as a function of time at four positions along the bilayer graphene device (blue, yellow, green and red). Dashed lines of slightly different color are solutions to Eq. \ref{eq:fick}  for $D^{\delta}=\unit[4\cdot10^{-5}]{cm^2/s}$ as explained in the text. Solutions for lower values of $D^{\delta}$ are shown as grey lines. Right panel: Li density profiles across the bilayer at different times (black lines), with measured values shown as colored dots. Grey lines show the density profile for the same lower $D^{\delta}$ values as in the left panel.}
	\label{fig:diff}
\end{figure}

\noindent
An elongated, narrow device geometry as shown at the top of Fig.~4 with multiple potential probes along the perimeter of uncovered bilayer graphene and with Li-ions, emanating from one side only, lends itself superbly to the study of the Li diffusion kinetics. As outlined previously, the density of Li-ions can be continuously monitored at discrete positions $x = x_{i}$ with $i = 0,1,2,\ldots$ as a function of time by simultaneously measuring the Hall voltage which develops across the bilayer graphene at these different distances from the graphene/electrolyte interface. In the device geometry depicted in Fig.~4, four Hall probe pairs are available, spaced $\unit[20]{\mu m}$ apart. As before, the density of Li-ions equals  $n_{\text{Li}}(x,t)=n_{e}(x,t)-n_{\text{imp}}(x,t=0)$, assuming a one-to-one charge transfer. The time dependence of diffusion-caused changes to the Li concentration is described by Fick's second law. For the one-dimensional case and an approximately constant $D^{\delta}$, this yields
\begin{equation}
\frac{\partial n_{\text{Li}}}{\partial t}=D^{\delta}\frac{\partial^2 n_{\text{Li}}}{\partial x^2}.
\label{eq:fick}
\end{equation}
This equation can be solved numerically (see supplementary information for details),
using the time-dependent density $n_{\text{Li}}(x=x_0=0,t)$ measured closest to the electrolyte as a boundary condition (due to this complex boundary condition in our experiment, there is no straightforward analytical solution). The finite length of the bilayer provides the second boundary condition: $\partial n_{\text{Li}}(x=x_{\rm end},t)/\partial x=0$. Here, $x_{\rm end}$ is the bilayer portion protruding furthest from the electrolyte at a distance $L$ from the first set of Hall probes at $x=x_0$. Here, $L$ equals $\unit[80]{\mu m}$. This leaves the diffusion coefficient $D^{\delta}$ as the only fit parameter.
\medskip

Li intercalation is triggered by the potentiostatic polarization of the device at $\unit[0.05]{V\text{ vs. Li/Li}^+}$. It is followed by fast diffusion of Li along the device. The density profiles $n_{\text{Li}}(x,t)$ at position $x_0 = 0, x_1 = 20\ {\rm \mu m}, x_2 = 40\ {\rm \mu m}\ \text{and}\ x_3 = 60\ {\rm \mu m}$ are plotted up to $\approx\unit[10^{13}]{cm^{-2}}$ in the left panel of Fig.~4. They have been extracted from Hall measurements as described above by subtracting the electron density due to initial doping from impurities at each position ($n_\text{imp}\approx\unit[2\cdot10^{12}]{cm^{-2}}$). They exhibit a steep increase with time. The Li density is always higher at positions closer to the electrolyte as expected. The profiles are numerically evaluated as described in the supplementary information; this has - in view of the very fast transport kinetics - the advantage that unavoidable preceding transients can be implicitly taken account of. The numerical solutions $n_{\text{Li}}(x,t)$ to Eq. \ref{eq:fick} for $x = x_1,\ x_2,\ \text{and}\ x_3$ are displayed as dashed lines (yellow, green, red, respectively) for $D^{\delta}=\unit[4\cdot10^{-5}]{cm^2/s}$. The experimentally extracted density profile at $x_0$ (blue) serves as a boundary condition for solving Eq. \ref{eq:fick} numerically. In order to illustrate how variations of the diffusion coefficient affect the density profiles, solutions for $n_{\text{Li}}(x_3,t)$ at a distance of $60\ {\rm \mu m}$ are also plotted for $D^{\delta}=\unit[1\cdot10^{-5}]{cm^2/s}$ and $D^{\delta}=\unit[1\cdot10^{-6}]{cm^2/s}$ (dark and light grey, respectively). These lower values of  $D^{\delta}$ lead to a clear discrepancy with experiment. In the right panel of Fig.~4, the density profiles along the device $n_{\text{Li}}(x,t)$ (black) are plotted at different times $t=\unit[0]{s}, \unit[3]{s}, \unit[6]{s}, \unit[9]{s}$ and $\unit[12]{s}$ for $D^{\delta}=\unit[4\cdot10^{-5}]{cm^2/s}$. For the sake of comparison, the experimentally measured densities have been superimposed. The striking agreement between the numerical solution and experiment corroborate further our extracted value of $D^{\delta}$ and the overall validity of our approach. Solutions $n_{\text{Li}}(x,t=\unit[12]{s})$ for $D^{\delta}=\unit[1\cdot10^{-5}]{cm^2/s}$ and $D^{\delta}=\unit[1\cdot10^{-6}]{cm^2/s}$ have been included in the right panel of Fig.~4 as well (dark and light grey lines, respectively) and confirm a significant deviation from experiment as $D^{\delta}$ is lowered from the best fit value.
\medskip

A closer inspection between the solution for $D^{\delta}=\unit[4\cdot10^{-5}]{cm^2/s}$ and the measurement reveals that within the data time frame displayed in Fig.~4, good agreement is achieved at intermediate densities while at both the lower and the upper density end the agreement is less convincing. The disagreement at the low density end, i.e., the case of very dilute intercalation, can be understood in terms of the small van-der-Waals gap between the graphene layers imposing a high activation energy for Li diffusion.\cite{Jungblut1989} This effectively implies the need for a reduction of the chemical diffusion coefficient $D^{\delta}$ to achieve a better match between the numerical solution and the measurement for this initial intercalation stage. With more Li entering in between the two graphene layers, the activation energy for diffusion decreases as the van-der-Waals gap is widened. $D^{\delta}$ consequently increases (for intermediate densities in Fig.~\ref{fig:diff}) until interactions between intercalated atoms make themselves felt by reducing it again. This density dependence as well as other influences making $D^{\delta}$ concentration dependent have not been taken account in Eq.~\ref{eq:fick}. Experimental and theoretical studies on graphitic carbon in the literature\cite{Yu1999,Persson2010} suggest a lowering of $D^{\delta}$ as $n_{\text{Li}}$ increases as the general trend, but its actual dependence bears a certain complexity as it involves possible competition between ordered intercalate phases. Our experimental findings would be in line with a decrease in $D^{\delta}$ at the high density end in Fig.~4. It is also important to note that the evaluated $D^{\delta}$ is distinctly higher than the surface diffusion coefficient\cite{Mandeltort2012}, thus ruling out intercalation to happen by surface diffusion followed by vertical intercalation.
\medskip

Apart from the above considerations, the kinetics is bound to also depend on the specific intercalation history.\cite{Magerl1990} During the first lithiation cycle, the solid-electrolyte interphase (SEI) is formed between bilayer graphene and the polymer electrolyte to establish a proper pathway for intercalation of Li into the bilayer. This is why we chose the 2nd lithiation cycle for the preceding analysis. Li diffusion may also be hampered by the presence of adsorbates or residues (e.g., from nanofabrication processes requiring resists), folds, ripples or other microscopic imperfections of the device. Some of these may be relieved at higher lithiation cycle numbers while others may potentially inflict a complete blocking of Li diffusion. The numerical solution to Eq.~\ref{eq:fick} following the above procedure suggests even higher chemical diffusion coefficients on the order of $D^{\delta}=\unit[7\cdot10^{-5}]{cm^2/s}$ during the 3rd and 4th lithiation cycle (see Fig.~S4 in the supplementary information). Intercalate diffusion during the 1st lithiation appears much slower ($D^{\delta}=\unit[5\cdot10^{-6}]{cm^2/s}$, see Fig.~S4) and reveals a discontinuity in the temporal evolution of the local Li density at one position. The latter is attributed to the sudden weakening of some local restraining condition.
\medskip

\begin{figure}
	\centering
	\includegraphics{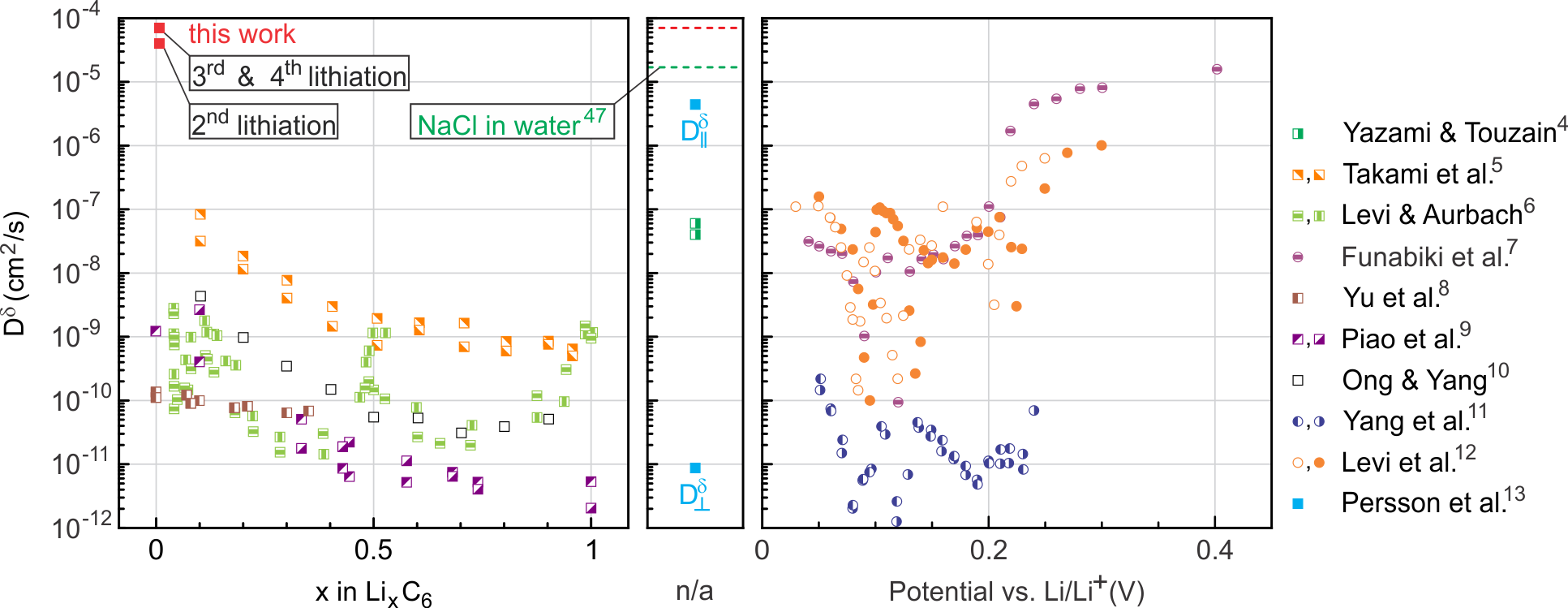}
	\caption{Room-temperature chemical diffusion coefficient $D^{\delta}$ of Li in graphite from different electrochemical experiments, reported as a function of composition $x$ in Li$_x$C$_6$ (left panel) or as a function of electrode potential vs. Li/Li$^+$ (right panel). In case neither is explicitly stated in the reference, data is shown in the middle panel together with $D^{\delta}$ for NaCl in water from Ref. \cite{Cussler1997} (green dashed line). See the respective references for details.  $D^{\delta}_{\perp}$ and $D^{\delta}_{\parallel}$ represent Li diffusion mainly perpendicular and parallel to the carbon sheets in bulk graphite, respectively. Values for $D^{\delta}$ of Li in bilayer graphene extracted from our experiment are shown as red squares.}
	\label{fig5}
\end{figure}

Fig.~5 compares the lithium in-plane diffusion coefficient obtained here on freely exposed bilayer graphene with experimental diffusion coefficient values reported in the literature for graphite. Data reported as a function of known or estimated values of lithium content $x$ in Li$_x$C$_6$ appear in the left graph, while data reported as a function of electrode potential vs. Li/Li$^+$ appear in the right graph. Experimental values for which neither $x$ nor the electrode potential is explicitly reported as well as $D^{\delta}$ for NaCl in water are given in the central graph. Data stem from electrochemical experiments near room temperature, which display a wide spread over seven orders of magnitude. The room temperature diffusion coefficients extracted in this work exceed all these values by one order of magnitude, suggesting a drastic enhancement of in-plane diffusion in bilayer graphene. In bulk graphite a comparably high diffusivity of Li has before only been observed in high-temperature studies, see Refs. \cite{Magerl1985, Jungblut1989}.
 
\bigskip

\noindent
\textbf{Conclusion}

\bigskip

\noindent
The implementation of a novel electrochemical cell architecture, consisting of a solidified polymer electrolyte confined at and covering only a small part of the perimeter of a patterned bilayer graphene device has enabled the operation of bilayer graphene as a true single phase mixed conductor. We have demonstrated gate-controlled reversible intercalation of Li-ions. In-situ time dependent Hall transport measurements reflect the charge density induced by the intercalant through charge transfer and provide unprecedented direct access to the intercalate diffusion kinetics due to the directionality of the device layout. Record in-plane chemical diffusion coefficients for lithium at room temperature up to $D^\delta=\unit[7\cdot10^{-5}]{cm^2/s}$ were disclosed. Intercalated Li-ions are also proven to enhance the intervalley scattering rate of charge carriers. We determine Li-ions to reside in between graphene sheets only, since diffusion in bilayer graphene stops at natural junctions with single layer graphene and characteristic Li-rich bilayer graphene edge decorations emerge at inner edges of macroscopic holes etched within the bilayer graphene device. The device concept with a local on-chip electrochemical cell implemented here is immediately transferable to other 2D materials as well as thin films. The marriage of electrochemical cell methods with magnetotransport tools commonly deployed in the physics of low dimensional electron systems offers access to powerful transport characterization methods previously not available in the context of ionic transport and diffusion. The immediate accessibility of the mixed conductor surface also offers the prospect of deploying local probe and surface analysis techniques to study the local kinetics and ordering of the intercalant otherwise hidden underneath the electrolyte.

\clearpage
\newpage

\noindent
\textbf{Methods}

\bigskip

\noindent
\textbf{Bilayer device fabrication\quad} Bilayer graphene flakes were obtained by mechanical exfoliation from bulk graphite (NGS Naturgraphit GmbH) using adhesive tape.\cite{Novoselov2004} Large bilayer flakes with typically at least one lateral dimension exceeding $\unit[50]{\mu m}$ were selected based on the optical contrast. The bilayer nature was additionally verified with Raman spectroscopy, in particular through the four-component line-shape of the 2D Raman band.\cite{Ferrari2006} A dry transfer technique\cite{Dean2010} was deployed to position and orient the flake on the target SiO$_2$-terminated silicon substrate. Electron beam lithography was used to pattern PMMA masks for shaping and/or isolating the flakes of interest via $\rm O_{\rm 2}$-plasma etching, as well as for the lift-off of metallic contacts to the bilayer flake and the counter-electrode contact for the electrochemical cell.\cite{Bonaccorso2012} All these contacts were made of $\unit[60]{nm}$-thick evaporated titanium. The bilayer contact configuration closely resembles a Hall bar geometry which is frequently exploited in two-dimensional electron systems for magnetotransport studies and the extraction of the electron density from the Hall effect. Completed devices were annealed in high-vacuum at $T>\unit[100]{\degree C}$ prior to depositing the electrolyte and running the electrochemical experiments.

\bigskip

\noindent
\textbf{Electrochemical cell fabrication\quad}
The fabrication of the electrochemical cell proceeded in an Ar-filled glovebox. The bilayer graphene device served as one electrode (working electrode, WE) of the cell. Metallic lithium was used as the counter electrode (CE). The lithium CE was realized by depositing metallic Li on a titanium contact pad. Only part of the bilayer graphene flake and the lithium counter electrode were covered with a polymer electrolyte composed of 0.35 M lithium bis(trifluoromethane) sulfonimide (LiTFSI) in polyethylene glycol methyl ether methacrylate (PEGMA):bisphenol A ethoxylate dimethacrylate (BEMA) w/w 3:7 with an added $\unit[2-4]{wt\%}$ of 2-hydroxy-2-methylpropiophenone, a common photoinitiator. After positioning, the electrolyte was solidified by UV-curing. More details about the electrolyte, elsewhere successfully employed for gating of few-layer graphene devices,\cite{Gonnelli2015} can be found in Ref.~\cite{Nair2011} as well as in the supplementary information.

\bigskip

\noindent
\textbf{Magnetotransport measurements\quad}
All magnetotransport measurements were performed in a cryogen-free superconducting magnet system with perpendicular magnetic fields $B$ up to $\unit[12]{T}$ and at temperatures between $\unit[1.4]{K}$ and $\unit[300]{K}$. Bilayer graphene devices were studied in vacuum with $p\le\unit[1\cdot10^{-6}]{mbar}$. Except for measurements of the chemical diffusion coefficient of Li, lock-in techniques were employed in a four-probe configuration to simultaneously measure the longitudinal and transverse voltage drops across the device. To this end, ac-excitation currents with typical rms-amplitudes $I\le\unit[1]{\mu A}$ at a frequency of $\unit[13.33]{Hz}$ were imposed through the device. For initial device characterization, we made use of the highly doped Si backgate, separated from bilayer graphene by a \unit[300]{nm}-thick thermally grown SiO$_2$. Assuming a simple parallel plate capacitor model with an areal capacitance $C/A=\epsilon_0\epsilon_r/d\approx\unit[11.5]{nF\cdot cm^{-2}}$, the electric field induced charge carrier density $\Delta n$ can be related to the applied backgate voltage $U_{BG}$ as $\Delta n/U_{BG}\approx\unit[7.2\cdot 10^{10}]{cm^{-2}V^{-1}}$. This model yields values typically in good agreement with those values extracted from measurements of the Hall coefficient $R_H=1/(qn)$ and Shubnikov-de Haas oscillations. Here, $q$ is the charge of the charge carriers. Prior to lithiation, the bilayer nature of our devices was confirmed by the characteristic integer Quantum Hall effect step of $8e^2/h$ in the transverse conductance $\sigma_{xy}$ around charge neutrality as the density was swept at low $T$ and a fixed high $B$-field.\cite{Novoselov2006} For measurements of the chemical diffusion coefficient of Li, a constant dc-current $I_{dc}\le\unit[1.5]{\mu A}$ was imposed across the bilayer graphene and a constant magnetic of $\unit[10]{T}$ was applied. Hall voltages were measured at different positions along the device using 7 1/2-digit PXI digital multimeters. The gate voltage between bilayer graphene and the lithium counter electrode was controlled at all times using a high-input impedance precision source/measure unit. Lithiation and delithiation is performed at room temperature at $\unit[0.05]{\text{V vs. Li/Li}^+}$  and $\unit[1.5]{\text{V vs. Li/Li}^+}$, respectively.

\clearpage

\begingroup
\renewcommand{\section}[2]{}%
\bibliography{references}
\endgroup

\bigskip

\noindent
\textbf{Acknowledgement\quad} We acknowledge financial support from the Baden-W\"urttemberg Stif\-tung as well as the EU graphene flagship and the DFG graphene priority programme (SPP 1459). The authors thank K.~v.~Klitzing for discussions and support, U.~Starke and T.~Acartürk for ToF-SIMS, and D.~Samuelis for initial discussions.

\bigskip

\noindent
\textbf{Author contributions\quad} M. K., F. P. and J. H. S. conceived the experiments. M. K. and F. P. fabricated the devices. M. K. performed the experiments. J. P. characterized the electrolyte. J. P. and J. M. contributed to the electrochemical design of the experiments. P. M. O. contributed the theoretical interpretation of the transport experiments. All Authors discussed the results. M. K. and J. H. S. wrote the manuscript and all authors contributed to it.

\bigskip

\noindent
\textbf{Additional Information\quad} Correspondence should be addressed to J. H. S.

\bigskip

\noindent
\textbf{Competing Interests\quad} The authors declare no competing financial interests.

\end{document}


\noindent
\Large\textbf{\textsf{Ultrafast lithium diffusion in bilayer graphene}}

\bigskip
\bigskip
\normalsize
\noindent
M. Kühne, F. Paolucci, J. Popovic, P. M. Ostrovsky, J. Maier \& J. H. Smet

\bigskip
\bigskip
\noindent
\textbf{Contents}
\bigskip

\tableofcontents

\bigskip
\bigskip
\section{Characterization of the solid polymer electrolyte}
The ionic conductivity of the poly(ethylene glycol) methyl ether methacrylate and bisphenol A ehtoxylate dimethacrylate (PEGMA, Mn = 500/BEMA, Mn = 1700) based solid polymer electrolytes prepared with different w/w ratios and different content of lithium bis(trifluoromethane) sulfonimide salt (LiTFSI) was measured upon UV curing by impedance spectroscopy in a self-made conductivity cell with stainless steel electrodes for a frequency range of $\unit[1-105]{Hz}$  at ambient temperature. The resulting impedance spectra (Fig.~S1) correspond to the bulk resistance of the polymer electrolytes yielding ionic conductivities ranging from $\sigma=\unit[2.8\cdot10^{-6}]{S/cm}$ (samples 3 and 4) to $\sigma=\unit[2.1\cdot10^{-5}]{S/cm}$ (sample 7). Based on the observed homogeneity and mechanical properties of the prepared materials, sample 5 (0.35 M LiTFSI in PEGMA/BEMA w/w 3:7 with 2-4 wt\% photoinitiator) with $\sigma=\unit[7.9\cdot10^{-6}]{S/cm}$ was selected as a suitable electrolyte for gate-controlled intercalation of our bilayer graphene devices.
\medskip

Besides ionic conductivity, the most important parameters for the lithium polymer electrolyte performance in a metallic lithium cell are the lithium transference number\cite{Evans1987} and the interfacial stability. The lithium transference number was measured by galvanostatic polarization in a self-made Li\textbar electrolyte\textbar Li cell similar as in Ref.~\cite{Popovic2016} (Fig.~S2). Upon applying a constant current $I$, the time dependent voltage change was recorded until the steady state was reached (Fig.~S2a). When the electrolyte does not react with lithium, i.e., the electrolyte resistance $R_{el}$ remains constant, the lithium transference number is determined from $t_{+}^{\rm pol}=\frac{IR_{tot,0}-IR_{SEI,0}}{U_\infty-IR_{SEI,\infty}}$, where $U_\infty$ is the steady state voltage. $R_{SEI,0}$ and  $R_{SEI,\infty}$ are resistances of the solid electrolyte interphase (SEI) before and after the polarization. The first semicircle before and after polarization in Fig.~S2b indicate a change to the electrolyte conductivity itself, which has to be taken into account ($\Delta R_{el}=R_{el,\infty}-R_{el,0}$). Hence, the transference number becomes $t_{+}^{\rm pol}=\frac{IR_{tot,0}-IR_{SEI,0}}{U_\infty-I(R_{SEI,\infty}+\Delta R_{el})}=0.40$. This Li-ion transference number being $<0.5$ can be expected for the given class of electrolyte material. Its finite value clearly demonstrates that the polymer electrolyte shows Li-ion conductivity.
\medskip

From Fig.~S2b it can be concluded that the reaction of $\unit[0.35]{M}$ LiTFSI in PEGMA/BEMA w/w 3:7 with 2-4 wt\% photoinitiator electrolyte with lithium both on the interface (SEI, second semicircle) and in the bulk (first semicircle) is not completed even after a 3 days aging process. While the SEI is observed to slightly grow with time, it is the bulk of the electrolyte which shows a more severe decrease in ionic conductivity. We attribute this behavior to the instability of the polymer electrolyte with respect to lithium. $\unit[0.35]{M}$ LiTFSI in PEGMA/BEMA w/w 3:7 with 2-4 wt\% photoinitiator nontheless proves a functional Li-ion conducting electrolyte whose mechanical properties make it particularly suitable for our Li-intercalation experiments.

\section{Numerical solution to Fick's second law}

We look for the numerical solution of Fick's second law in one dimension (Eq. 2 in the main text)
\begin{equation}
\frac{\partial n_\text{Li}(x,t)}{\partial t}=D^{\delta}\frac{\partial^2 n_\text{Li}(x,t)}{\partial x^2},
\label{eq:Sfick}
\end{equation}
with the following boundary conditions
\begin{align}
 n_{\text{Li}}(0,t) & =  n^\text{meas}_\text{Li}(0,t) &  \left.\frac{\partial n_\text{Li}(x,t)}{\partial x}\right|_{x=L}=0.
 \label{eq:Sbound}
\end{align}
The first boundary condition sets the lithium density at $x=0$ equal to the measured one at all times $t$. The second boundary condition imposes zero flux at the ionically blocking electrode at $x=L$ at all times $t$. We replace the derivatives in time and space by their approximative finite-difference formulations
\begin{align}
\left.\frac{\partial n_\text{Li}}{\partial t}\right|_{x_i,t_j} & =\frac{n_i^{j+1}-n_i^j}{\Delta t}, &
\left.\frac{\partial^2 n_\text{Li}}{\partial x^2}\right|_{x_i,t_j} & =\frac{n_{i+1}^j-2n_i^j+n_{i-1}^j}{(\Delta x)^2}.
\label{eq:Sdiff}
\end{align}
Here we make the transition from continuous variables $x$ and $t$ to discrete ones $x_i$ and $t_j$, and we write $n_i^j=n_\text{Li}(x_i,t_j)$, $\Delta t=t_{j+1}-t_j$ and $\Delta x=x_{i+1}-x_i$. Combining (\ref{eq:Sfick}) and (\ref{eq:Sdiff}), we obtain
\begin{equation}
n_i^{j+1}=n_i^j+D^{\delta}\Delta t\frac{n_{i+1}^j-2n_i^j+n_{i-1}^j}{(\Delta x)^2},
\label{eq:Scalc}
\end{equation}
which gives a solution stable in time provided
\begin{equation}
\frac{D^{\delta}\Delta t}{(\Delta x)^2}\le \frac{1}{2}.
\label{eq:Sstab}
\end{equation}
As shown in Fig.~S3, we divide the region of interest of the sample of length $L$ into $N_x=L/\Delta x+1$ discrete steps in $x$ direction. In our case (device shown in Fig.~4 in the main text) $L=\unit[80]{\mu m}$ and $\Delta x=\unit[1]{\mu m}$ yielding $N_x=81$ steps. We discretize the time frame of interest into $N_t$ steps depending on the choice of $D^{\delta}$ respecting the stability condition (\ref{eq:Sstab}). We start with an initial density profile $n_i^{0}=0$ for all $i=0\dots(N_x-1)$. Taking into account the boundary conditions (\ref{eq:Sbound}), we first set $n_{0}^{1}=n_\text{Li}^\text{meas}(0,t_1)$, then calculate (\ref{eq:Scalc}) for $i=1\dots (N_x-2)$ and finally set $n_{i=N_x-1}^{1}=n_{i=N_x-2}^{1}$. We then proceed with $t_j$ for $j=2\dots (N_t-1)$. Depending on the choice of $D^{\delta}$, $\Delta t$ needs to be chosen such that (\ref{eq:Sstab}) is fulfilled. Hence, $n_{0}^j$ is extracted from the measured $n_\text{Li}(0,t)$ by interpolation, if needed. The solution provides us with both Li density profiles across the sample at a given time $t_j$, as well as the time-dependent Li density evolutions at a specific position $x_i$.

\section{Li diffusion}

In Fig.~S4, the time-dependence of the Li density $n_{\text{Li}}(x,t)$ extracted from Hall measurements of the bilayer graphene device illustrated in Fig.~4 of the main text are displayed for the 1$^{\text{st}}$, 3$^{\text{rd}}$ and 4$^{\text{th}}$ lithiation. Again the measured data (colored, solid lines) corresponds to simultaneous measurements taken at the same four positions $x$ outside the electrolyte and spaced apart by $\unit[20]{\mu m}$ as described in the main text.
\medskip

For the 3rd lithiation, good agreement with a numerical solution to Fick's second law (\ref{eq:Sfick}) can be obtained for $D^{\delta}\approx\unit[7\cdot10^{-5}]{cm^2/s}$. For the 4th lithiation, this value even seems to underestimate the fast Li diffusion measured in the experiment. As discussed in the main text, Li diffusion is slower during the very first lithiation (top panel in Fig.~S4). Here, $n_\text{Li}^\text{meas}(x=\unit[40]{\mu m},t)$ (green solid line) also shows unexpected behavior which may be attributed to the sudden loosening of a local restraining condition on Li accumulation at $t\approx\unit[230]{s}$ (such as residues located on top of the bilayer or possibly a wrinkle, locally blocking Li atoms from entering a certain region). After the 1$^{\text{st}}$ lithiation, intercalate diffusion seems to be more reproducible with a tendency to become more facile at higher lithiation cycle numbers.

\bigskip
\bigskip
\begingroup
\renewcommand{\section}[2]{}%
\bibliography{references}
\endgroup

\clearpage
\begin{figure}
	\centering
	\includegraphics{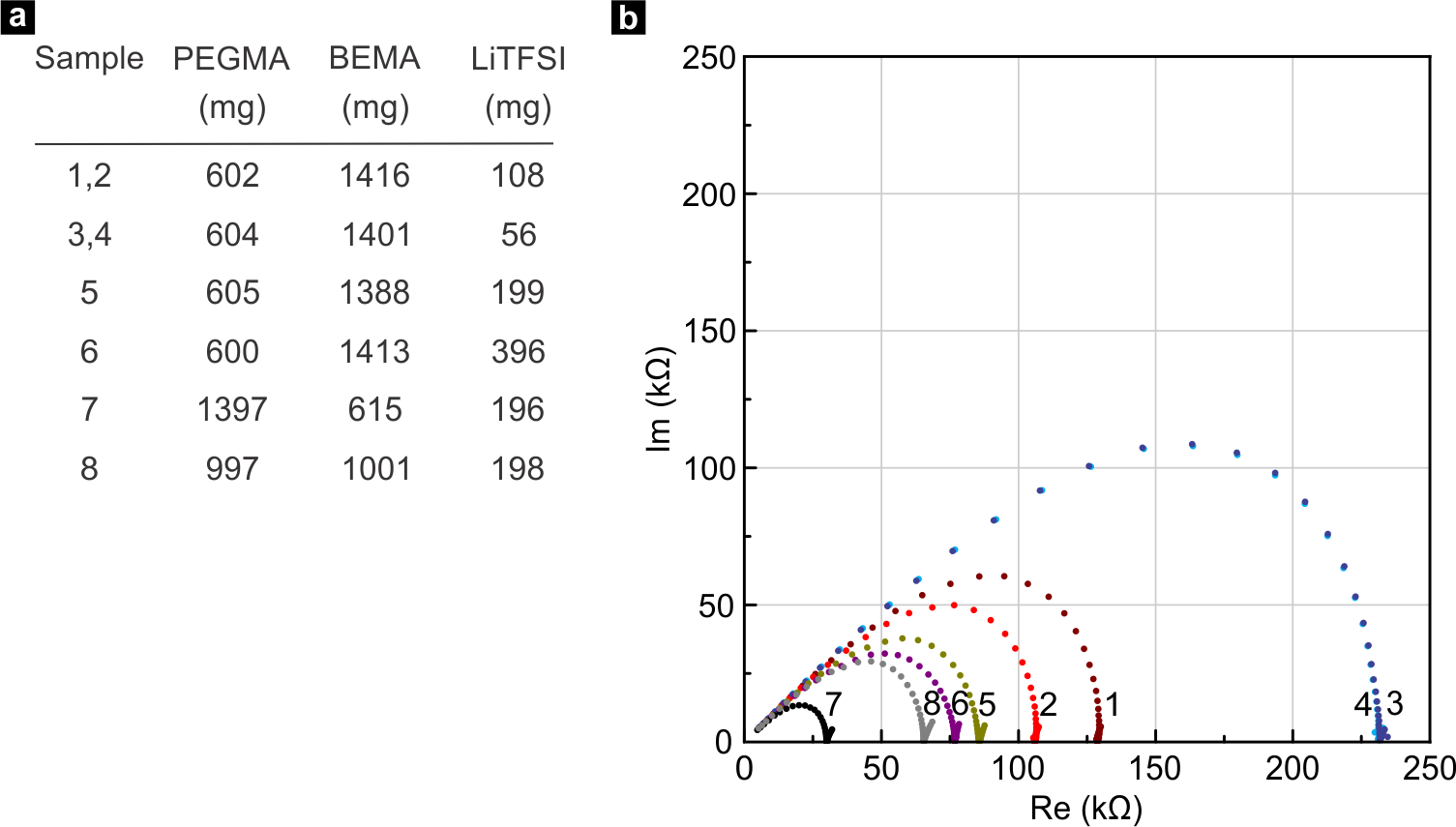}
	\caption{Characterization of eight different LiTFSI in PEGMA/BEMA solidified polymer electrolyte samples with 2-4 wt\% photoinitiator at ambient temperature. (a) The composition of the studied samples. (b) Impedance spectroscopy of all samples.}
	\label{fig:Imp}
\end{figure}

\clearpage
\begin{figure}
	\centering
	\includegraphics{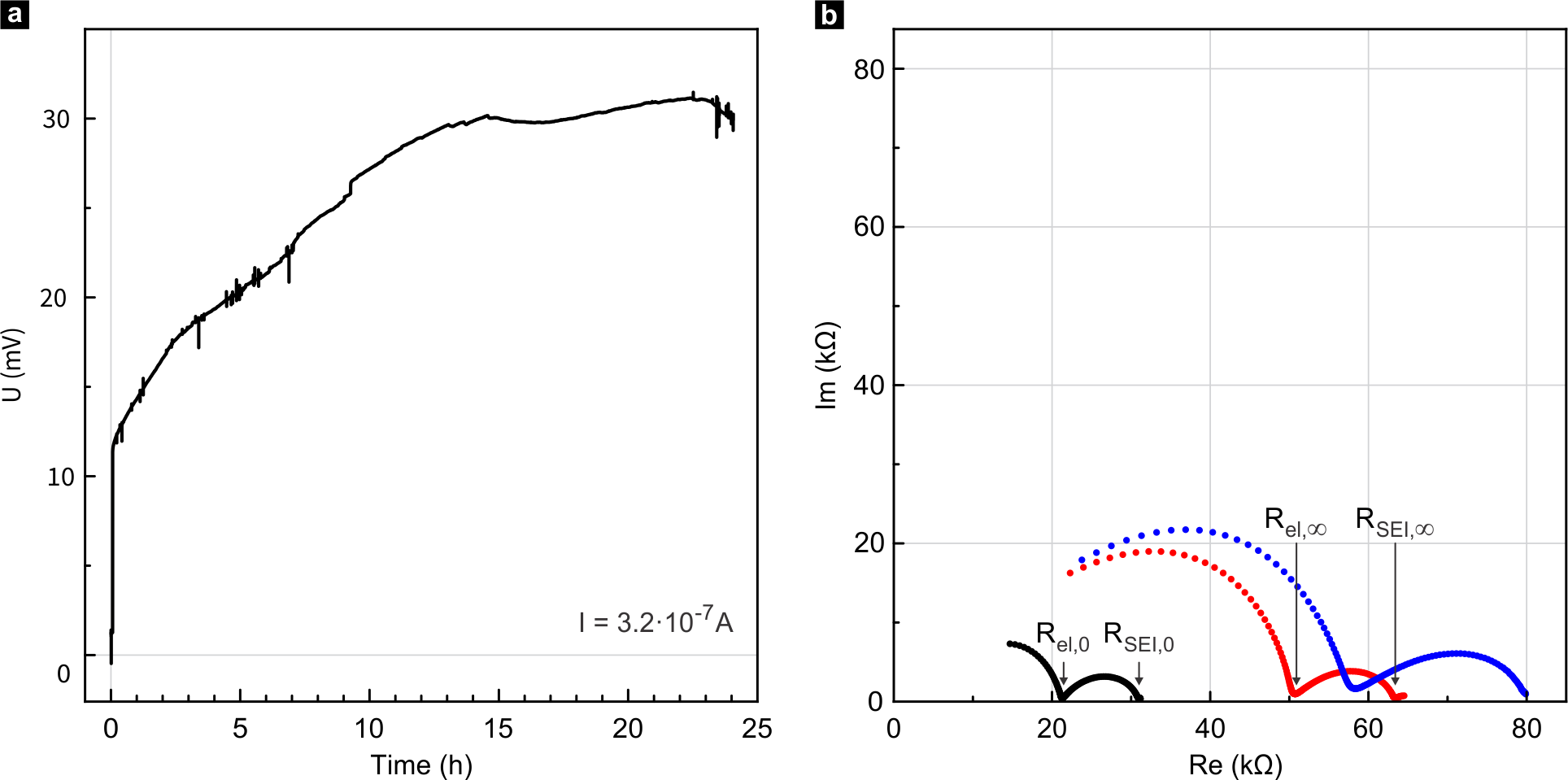}
	\caption{(a) Galvanostatic polarization of $\unit[0.35]{M}$ LiTFSI in PEGMA/BEMA w/w 3:7 with 2-4 wt\% photoinitiator, (b) Impedance spectra before polarization (black), after polarization (red) and 3 days after polarization (blue). All measurements are performed in a Li\textbar electrolyte\textbar Li cell at room temperature.}
	\label{fig:trans}
\end{figure}

\clearpage
\begin{figure}
	\centering
	\includegraphics{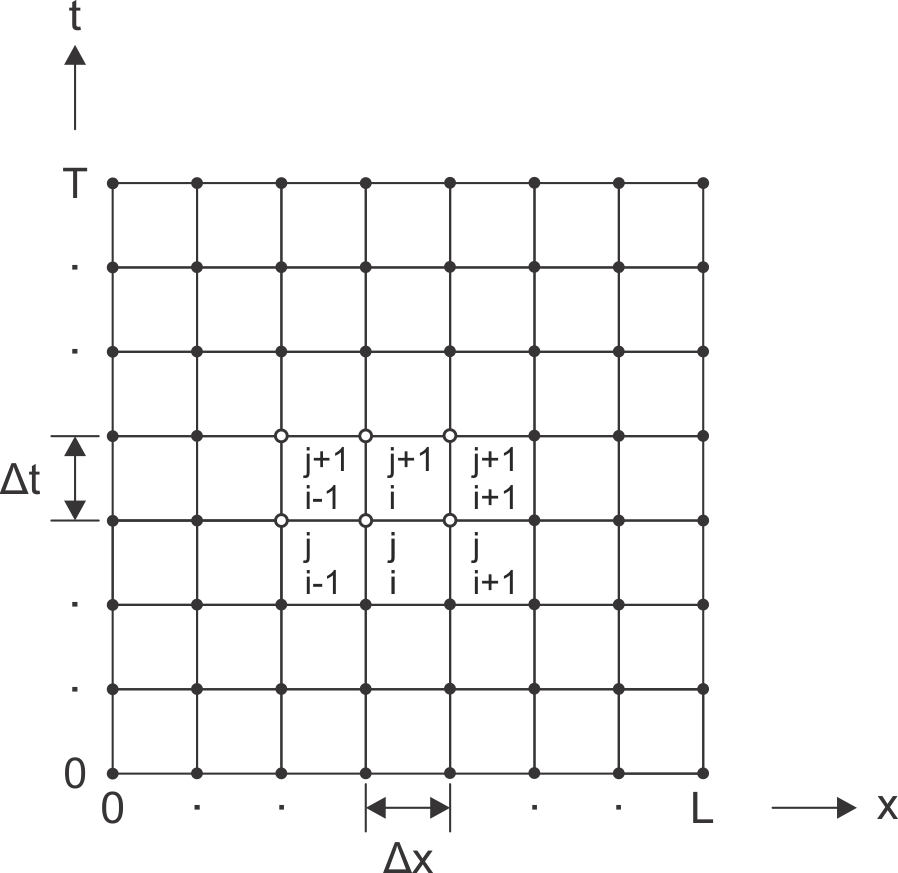}
	\caption{Schematic grid used for our numerical calculation of Fick's second law.}
	\label{figS1:num}
\end{figure}

\clearpage
\begin{figure}
	\centering
	\includegraphics{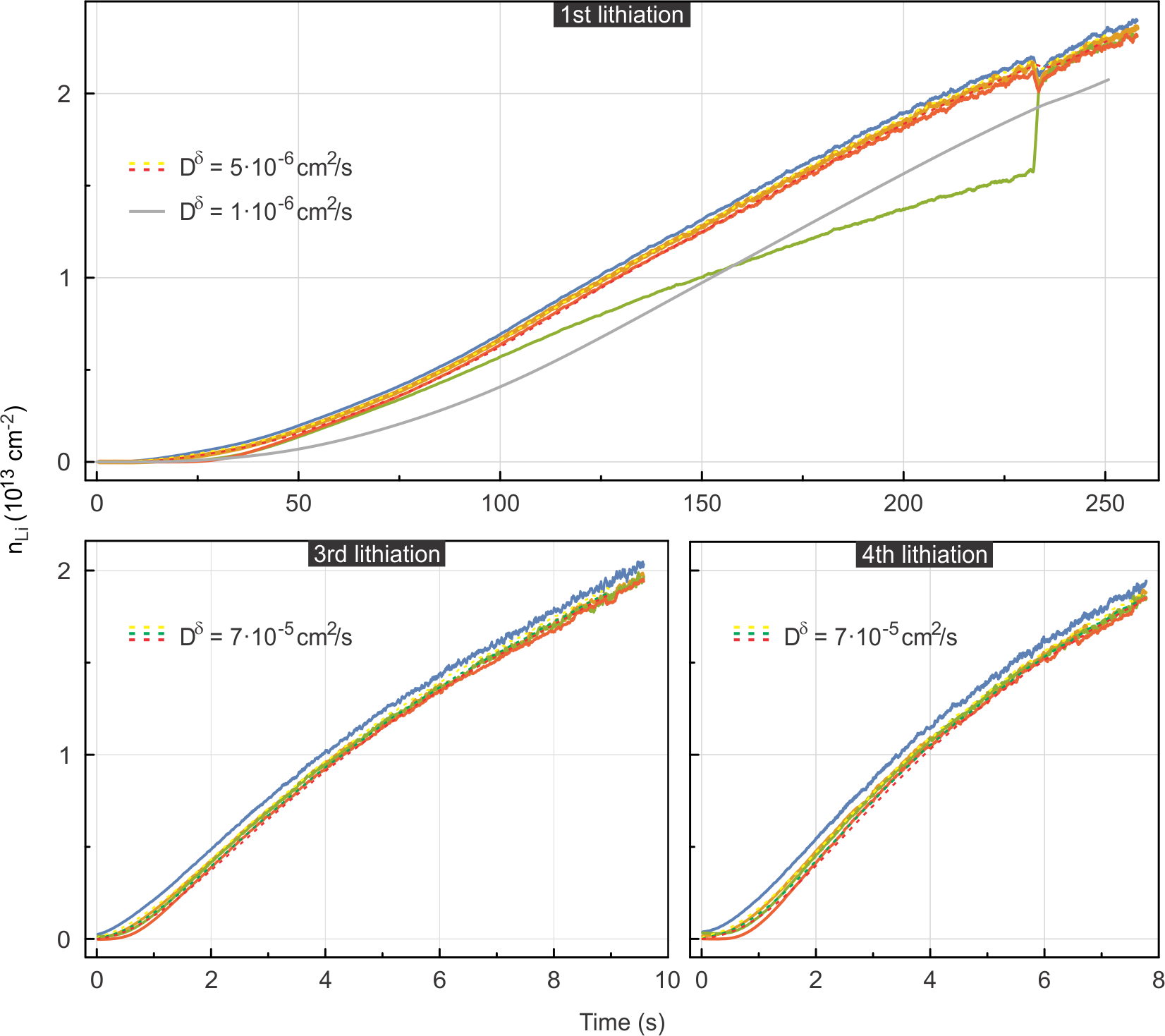}
	\caption{Li diffusion in the bilayer graphene device shown in Fig.~4 of the main text: 1st, 3rd \& 4th lithiation. The yellow, green and red dashed lines are numerical fits to data (solid lines) measured at $x=\unit[20]{\mu m}$, $\unit[40]{\mu m}$ and $\unit[60]{\mu m}$, respectively. In the upper panel, $n_\text{Li}^\text{meas}(x=\unit[40]{\mu m},t)$ shows an unexpected time dependence, presumably due to a local restraint on Li accumulation suddenly loosening at $t\approx\unit[230]{s}$. $n_\text{Li}(x=\unit[60]{\mu m},t)$ calculated for $D^{\delta}=\unit[1\cdot10^{-6}]{cm^2/s}$ is plotted as a solid grey line.}
	\label{figS:dif}
\end{figure}